\documentclass[11pt]{article}
\usepackage{graphicx}
\topmargin - 2cm
\oddsidemargin - 0.1cm
\evensidemargin - 5cm
\newcommand{\eproof}{\rule{0.2cm}{0.2cm}}

\newtheorem{thm}{Theorem}[section]
\newtheorem{prop}[thm]{Proposition}
\newtheorem{lem}[thm]{Lemma}
\newtheorem{cor}[thm]{Corollary}
\newtheorem{definition}[thm]{Definition}
\newtheorem{remark}[thm]{Remark}

\begin{document}

\title{\small{{\bf  SYMMETRIC $(q,\alpha)$-STABLE DISTRIBUTIONS.  \\
PART I: FIRST REPRESENTATION}}}
\author{Sabir Umarov$^{1}$, Constantino Tsallis$^{2,3}$, Murray Gell-Mann$^{3}$ \\and Stanly Steinberg$^{4}$}
\date{}
\maketitle
\begin{center}
$^{1}$ {\it Department of Mathematics\\
Tufts University, Medford
MA 02155, USA}\\

$^2$ {\it Centro Brasileiro de Pesquisas Fisicas \\
Xavier Sigaud 150, 22290-180 Rio de Janeiro-RJ, Brazil}

$^{3}$ {\it Santa Fe Institute \\
1399 Hyde Park Road, Santa Fe, NM 87501,
USA}\\

$^{4}$ {\it Department of Mathematics and Statistics\\
University of New Mexico,
Albuquerque, NM 87131, USA}\\

\end{center}

%%%%%%%%%%%%%%%%%%%%%%%%%%%%%%%%%%%%%%%%%%%%%%%%%%%%%%%%%%%%%%%%%%%%%%%%%%%%

\begin{abstract}
The classic central limit theorem and $\alpha$-stable distributions
play a key role in probability theory, and also in Boltzmann-Gibbs
(BG) statistical mechanics. They  both concern the paradigmatic case
of probabilistic independence of the random variables that are being
summed. A generalization of the BG theory, usually referred to as
nonextensive statistical mechanics and characterized by the index
$q$ ($q=1$ recovers the BG theory), introduces special (long range)
correlations between the random variables, and recovers independence
for $q=1$. Recently, a $q$-central limit theorem consistent with
nonextensive statistical mechanics was established
\cite{UmarovTsallisSteinberg} which generalizes the classic Central
Limit Theorem. In the present paper we introduce and study symmetric
$(q,\alpha)$-stable distributions. The case $q=1$ recovers the
L\'evy $\alpha$-stable distributions.
\end{abstract}
%%%%%%%%%%%%%%%%%%%%%%%%%%%%%%%%%%%%%%%INTRODUCTION
\section{Introduction}
In paper \cite{UmarovTsallisSteinberg} a generalization of the
classic central limit theorem applicable to nonextensive statistical
mechanics \cite{Tsallis1988,PratoTsallis1999} (which recovers the
usual, Boltzmann-Gibbs statistical mechanics as the $q=1$ particular
instance), was presented; for reviews of nonextensive statistical
mechanics see \cite{GellmannTsallis,BoonTsallis,encyclopedia}. We
follow here along the lines of that paper. One of the important
aspects of this generalization is that it concerns the case of
random variables correlated in a special manner. On the basis of the
$q$-Fourier transform $F_q$ introduced there ($F_1$ being the
classical Fourier transform), and the function
$$
z(s)=\frac{1+s}{3-s} \,,
$$
we described attractors of conveniently scaled limits of sums of
$q$-independent random variables \footnote{$q$-independence
corresponds to standard probabilistic independence if $q=1$, and to
specific long-range (global in physics terminology) correlations if
$q \ne 1$. } with a finite $(2q-1)$-variance \footnote{We required
there $1 \le q<2$. Denoting $Q=2q-1$, it is easy to see that this
condition is equivalent to the finiteness of the $Q$-variance with
$1\le Q<3$; see also \cite{TsallisPlastinoAlvarez}.}. This description was essentially based on the mapping
\begin{equation}
\label{paper1}
F_q: \mathcal{G}_q [2] \rightarrow \mathcal{G}_{z(q)}[2] \,,
\end{equation}
where $\mathcal{G}_q[2]$ is the set of $q$-Gaussians (the number 2 in the notation will soon become transparent).

In the current work, which consists of two parts, we will introduce
and study a $q$-analog of the $\alpha$-stable L\'evy distributions.
In this sense, the present paper is a conceptual continuation of
paper \cite{UmarovTsallisSteinberg}. For  simplicity we will analyze
only symmetric densities in the one-dimensional case. The classic
theory of $\alpha$-stable distributions $\mathcal{L}(\alpha)$ was
originated by Paul L\'evy and developed by  L\'evy, Gnedenko, Feller
and others (see, for instance,
\cite{KolmogorovGnedenko,Feller,SamorodnitskyTaqqu,UchaykinZolotarev,MeerschaertScheffler}
and references therein for details and history). The $\alpha$-stable
distributions found a huge number of applications in various
practical studies
\cite{Zaslavsky,Beckbook,MetzlerKlafter,GorenfloMainardi,MeerschaertScheffler1,SchmittSeuront,variousCLT,TsallisBukman,AbeRajagopal2000,Tsallis2005,MoyanoTsallisGellmann2006},
confirming the frequent nature of these distributions.

We introduce a class of random variables $\mathcal{L}_q(\alpha),$
which we call $(q,\alpha)$-stable distributions. Namely, we consider
the symmetric densities $f(x)$ with asymptotics $f \sim C
|x|^{-\frac{1+\alpha}{1+\alpha(q-1)}}, \, \, |x|\rightarrow \infty$,
where $1\le q <2, ~ 0<\alpha < 2,$ and $C$ is a positive constant
\footnote{Hereafter $g(x) \sim h(x), x \rightarrow a,$ means that
$\lim_{x \rightarrow a}\frac{g(x)}{h(x)} = 1$.}. We establish that
linear combinations and scaling limits of sequences of
$q$-independent random variables with $(q,\alpha)$-stable
distributions are again random variables with $(q,\alpha)$-stable
distributions. These facts justify that $\mathcal{L}_q(\alpha)$ form
a class of stable distributions. To this end, we note that
$\mathcal{L}_q(\alpha)$ for fixed $q \in [1,2)$ and $\alpha \in
(0,2)$ coincides with the set of symmetric L\'evy distributions
$\mathcal{L}_{sym}(\gamma),$ where
$$\gamma=\gamma(q,\alpha)=\frac{\alpha (2-q)}{1+\alpha(q-1)}.$$
However, the $(q,\alpha)$-stability holds for specifically
correlated random variables ($q$-independence exhibits a special
correlation). Thus for each fixed $q\in[1,2)$ we have the set of
$q$-independent $(q,\alpha)$-stable distributions. If $q=1$ then
$q$-independence becomes usual independence and
$\gamma(1,\alpha)=\alpha,$ implying
$\mathcal{L}_1(\alpha)=\mathcal{L}_{sym}(\alpha)$.  The main purpose
of the current paper is to classify $(q,\alpha)$-stable
distributions in terms of their densities depending on the
parameters $1 \le q < 2$ (or equivalently $1 \le Q<3$, $Q=2q-1$) and
$0<\alpha\leq 2$. We establish the mapping
\begin{equation}
\label{paper2part1}
F_q: \mathcal{G}_{q^{L}} [2] \rightarrow \mathcal{G}_{q}[\alpha],
\end{equation}
where $\mathcal{G}_{q}[\alpha]$ is the set of functions
$\{be_q^{-\beta |\xi|^{\alpha}}, \, b>0, \, \beta >0 \},$ and
$$
q^{L}=\frac{3+Q\alpha}{1+ \alpha}, \, Q=2q-1 \,,
$$
i.e.,
$$
\frac{2}{q^L-1}=\frac{1+\alpha}{1+\alpha(q-1)} \,.
$$
The particular case $q=Q=1$ recovers $q^L=\frac{3+\alpha}{1+ \alpha}$, already known in the literature \cite{PratoTsallis1999}.
%We consider the values of parameters $Q$ and $\alpha$ ranging in the set
Denote $ \mathcal{Q}_1=\{(Q,\alpha): 1 \le Q < 3, \, \alpha = 2 \},$
$ \mathcal{Q}_2=\{(Q,\alpha): 1 \le  Q < 3, \, 0<\alpha < 2 \}$ and
$\mathcal{Q} = \mathcal{Q}_1 \cup \mathcal{Q}_2.$ Note that the case
$(Q,\alpha) \in \mathcal{Q}_1$ for $q$-independent random variables
with a finite $Q$-variance was studied in
\cite{UmarovTsallisSteinberg}.
For $(Q,\alpha) \in \mathcal{Q}_2$ the $Q$-variance is infinite. We
will focus our analysis namely on the latter case. Note that the
case $\alpha=2$, in the framework of the present description like in
that of the classic $\alpha$-stable distributions, becomes peculiar.

In Part II we study the attractors of scaled sums, and expand the
results of paper \cite{UmarovTsallisSteinberg} to the region
$\mathcal{Q}$ generalizing the mapping (\ref{paper1}) into the form
\begin{equation}
\label{paper2part2} F_{\zeta_{\alpha}(q)}: \mathcal{G}_q [\alpha]
\rightarrow \mathcal{G}_{z_{\alpha}(q)}[\alpha], \, 1 \le q<2, \,
0<\alpha \leq 2,
\end{equation}
where
\[
\zeta_{\alpha}(s) = \frac{\alpha - 2(1-q)}{\alpha} \, \, \mbox{and} \,
\, z_{\alpha}(s)=\frac{\alpha q + 1 - q}{\alpha + 1 - q.}
\]
Note that, if $\alpha = 2$, then $\zeta_{2}(q)=q$ and
$z_2(q)=(1+q)/(3-q),$ thus recovering the mapping (\ref{paper1}),
and consequently, the result of \cite{UmarovTsallisSteinberg}.

These two descriptions of $(q,\alpha)$-stable distributions, based
on mappings (\ref{paper2part1}) and (\ref{paper2part2})
respectively, can be unified to the scheme
\begin{equation}
\label{schemeLevy3} \mathcal{L}(q,\alpha)
\stackrel{F_{q}}{\longrightarrow} \mathcal{G}_q(\alpha)
\stackrel{F_{q_{\zeta(\alpha)}}}{\longleftrightarrow}
\mathcal{G}_{q_{\zeta(\alpha)}}(2)
\end{equation}
\hspace{3.2in} $\updownarrow \, F_{q}$
\[
\hspace{0.1in} \mathcal{G}_{q^{L}} (2),
\]
which gives the full picture of interrelations between the values of
parameters $q \in [1,2)$ and $\alpha \in(0,2).$

\section{Basic operations of $q$-algebra}
We recall briefly the basic operations of $q$-algebra. Indeed, the
analysis we will conduct is entirely based on the $q$-structure of
nonextensive statistical mechanics (for more details see
\cite{TsallisQuimicaNova,GellmannTsallis}
%,qborges1,qnivanen,qborges2}
and references therein). To this end, we recall the well known fact
that the classical Boltzmann-Gibbs entropy $S_{BG}=-\sum_i p_i \ln
p_i$ satisfies the additivity property. Namely, if $A$ and $B$ are
two independent subsystems, then
$
\label{additivity} S_{BG}(A+B)=S_{BG}(A)+S_{BG}(B).
$
However, the $q$-generalization of the classic entropy introduced in
\cite{Tsallis1988} and given by $S_q=\frac{1-\sum_i p_i^q}{q-1}$
with $q\in \cal{R}$ and $S_1 = S_{BG}$, does not possess this
property if $q \neq 1$. Instead, it satisfies the {\it
pseudo-additivity} (or {\it $q$-additivity})
\cite{Tsallis1988,PratoTsallis1999,Tsallis2005}
\[
S_q(A+B)=S_q(A)+S_q(B)+(1-q)\,S_q(A)\,S_q(B).
\]
Inherited from the right hand side of this equality, the {\it
$q$-sum} of two given real numbers, $x$ and $y$, is defined as $x
\oplus_q y = x+y+(1-q)xy$. The $q$-sum is commutative, associative,
recovers the usual summing operation if $q=1$ (i.e. $x \oplus_1 y =
x+y$), and preserves $0$ as the neutral element (i.e. $x \oplus_q 0
= x$). By inversion,
we can define the {\it $q$-subtraction} as $x \ominus_q y =
\frac{x-y}{1+(1-q)y}.$ The {\it $q$-product} for $x,y$ is defined by
the binary relation $x \otimes_q y =
[x^{1-q}+y^{1-q}-1]_+^{{1}\over{1-q}}.$ Here the symbol $[x]_+$
means that $[x]_+ = x$ if $x \geq 0$, and $[x]_+ = 0$ if $x < 0.$
This operation also commutative, associative, recovers the usual
product when $q=1$, and preserves $1$ as the unity. The $q$-product
is defined if $x^{1-q}+y^{1-q} \geq 1$. Again by inversion, it can
be defined the {\it $q$-division}: $x \oslash_q y =
(x^{1-q}-y^{1-q}+1)^{1 \over {1-q}}.$

\section{$q$-generalization of the exponential and cyclic functions}
\par
Now we introduce the $q$-exponential and $q$-logarithm
\cite{TsallisQuimicaNova}, which play an important role in the
nonextensive theory. These functions are denoted by $e_q^x$ and
$ln_q x$ and respectively defined as $e_q^x=[1+(1-q)x]_+^{{1}\over
{1-q}}$ and $\ln_q x= \frac{x^{1-q}-1}{1-q}, \, (x>0).$ The entropy
$S_q$ can be conveniently rewritten in the form $S_q=\sum_i p_i
\ln_q {1 \over p_i}.$

Let us mention now the main properties of these functions, which we
will use essentially in this paper. For the $q$-exponential the
relations $e_q^{x \oplus_q y} = e_q^x e_q^y$ and $e_q^{x+y}=e_q^x
\otimes_q e_q^y$ hold true. These relations can be written
equivalently as follows: $\ln_q (x \otimes_q y)=\ln_q x + \ln_q y$
\footnote{This property reflects the possible extensivity of $S_q$
in the presence of special correlations
\cite{TsallisGellmannSato,MarshallEarl,tsallisEPN,MarshFuentesMoyanoTsallis}.},
and $\ln_q (x y)=(\ln_q x) \oplus_q (\ln_q y)$. The $q$-exponential
and $q$-logarithm have the asymptotics
\begin{equation}
\label{exp}
e_q^x = 1 + x + {q \over 2}x^2 + o(x^2), \, x \rightarrow 0,
\end{equation}
and
\begin{equation}
\label{log}
\ln_q (1+x) = x - {q \over 2} x^2 + o(x^2), \, x \rightarrow 0,
\end{equation}
respectively. The $q$-product and $q$-exponential can be extended to
complex numbers $z=x+iy$ (see \cite{UmarovTsallisSteinberg,QT,QT2}).

In addition, for $q \neq 1$ the function $e_q^z$ can be analytically
extended to the complex plain except the point $z_0=-1/(1-q)$ and
defined as the principal value along the cut $(-\infty,z_0).$  If $q
< 1,$ then, for real $y$, $|e_q^{iy}| \geq 1$ and $|e_q^{iy}| \sim
K_q (1+y^2)^{\frac{1}{2(1-q)}}, \, y \rightarrow \infty,$ with
$K_q=(1-q)^{1/(1-q)}.$ Similarly, if $q > 1$, then $0 < |e_q^{iy}|
\leq 1$ and $|e_q^{iy}| \rightarrow 0$ if $|y| \rightarrow \infty.$
\begin{lem}
\label{psr} Let $A_n(q)= \prod_{k=0}^{n} a_k(q),$ where
$a_k(q)=q-k(1-q).$ Then the following power series expansion holds
\[
e_q^z = 1 +z + z^2 \sum_{n=0}^{\infty} \frac{A_n(q)}{(n+2)!} z^n, \,
\, |z|< \frac{1}{|1-q|}.
\]
\end{lem}
\begin{cor} Let $I_q = (-1/|1-q|, 1/|1-q|).$ For arbitrary real number $x\in I_q$ the equation
\[
e_q^{ix} = \{ 1 - x^2 \sum_{n=0}^{\infty} \frac{(-1)^n
A_{2n}(q)}{(2n+2)!} x^{2n} \} +
               i \{  x - x^2 \sum_{n=0}^{\infty} \frac{(-1)^n
A_{2n+1}(q)}{(2n+3)!} x^{2n+1}  \}
\]
holds.
\end{cor}
Define for $x \in I_q$ the functions $q$-cos and $q$-sin by formulas
\begin{equation}
\label{cos}
\cos_q(x) = 1 - x^2 \sum_{n=0}^{\infty} \frac{(-1)^n
A_{2n}(q)}{(2n+2)!} x^{2n},
\end{equation}
and
\begin{equation}
\label{sin}
\sin_q (x) = x - x^2 \sum_{n=0}^{\infty} \frac{(-1)^n
A_{2n+1}(q)}{(2n+3)!} x^{2n+1}.
\end{equation}
In fact, $\cos_q(x)$ and $\sin_q(x)$ is defined for all real $x$ by
using appropriate power series expansions. Properties of $q$-sin,
$q$-cos, and corresponding $q$-hyperbolic functions, were studied in
\cite{qborges1}. Here we note that the $q$-analogs of the well known
Euler's formulas read
\begin{cor}

\begin{enumerate}
\item[(i)]
$
e_q^{ix}=\cos_q(x)+i \, \, \sin_q(x);
$
\item[(ii)]
$\cos_q(x)=\frac{e_q^{ix}+e_q^{-ix}}{2};$
\item[(iii)]
$\sin_q(x)=\frac{e_q^{ix}-e_q^{-ix}}{2i}.$

\end{enumerate}
\end{cor}
\begin{lem}
The following equality holds:
\begin{equation}
\label{cos2x}
\cos_q (2x) = e_{2q-1}^{2(1-q)x^2} -2 \,  \sin_{2q-1}^2 (x).
\end{equation}
\end{lem}
\vspace{.3cm}

{\it Proof.} The proof follows from the definitions of $\cos_q (x)$
and $\sin_q (x),$ and from the fact that $(e_q^{x})^2 =
e_{(1+q)/2}^{2x}$ (see Lemma 2.1 in \cite{UmarovTsallisSteinberg}).
\eproof \vspace{.3cm}

Denote $\Psi_q(x)=\cos_q 2x -1.$ It follows from Equation (\ref{cos2x}) that
\begin{equation}
\label{psi}
\Psi_{q}(x)= ( e_{2q-1}^{2(1-q)x^2} - 1 ) -2 \,  \sin_{2q-1}^2 (x).
\end{equation}
\begin{lem}
\label{lempsi}
Let $q \ge 1$. Then we have
\begin{enumerate}
\item[1.]
$-2 \le \Psi_q(x) \le 0;$
\item[2.]
$\Psi_q(x) = - 2 \, q \, x^2 + o(x^3), \, x \rightarrow 0.$
\end{enumerate}
\end{lem}
\vspace{.3cm}

{\it Proof.}
It follows from (\ref{psi}) that $\Psi_q(x) \le 0.$ Further, $\sin_{q}(x)$ can be written in the form (see \cite{qborges1})
$\sin_q(x)=\rho_q(x) \, \sin[\varphi_q(x)],$ where $\rho_q(x)=(e_q^{(1-q)x^2})^{1/2}$ and $\varphi_q(x)=\frac{\arctan(1-q)x}{1-q}.$
This yields $\Psi_q (x) \ge -2$ if $q \ge 1.$ Using the asymptotic relation (\ref{exp}), we get
\begin{equation}
\label{one}
e_{2q-1}^{2(1-q)x^2} - 1 = 2(1-q)x^2 +o(x^3), \, x \rightarrow 0.
\end{equation}
It follows from (\ref{sin}) that
\begin{equation}
\label{two}
-2 \, \sin_{2q-1}^2 (x) = -2 \, x^2 + o(x^3), x\rightarrow 0.
\end{equation}
The relations (\ref{psi}), (\ref{one}) and (\ref{two}) imply the
second part of the statement. \eproof \vspace{.3cm}

\begin{remark}
It is not hard to verify that in the case $q>1$ for $x>(q-1)^{-1}$
the representation
\[
e_q^{-x}=[(q-1)x]^{-\frac{1}{q-1}}\left(1-\frac{1}{(1-q)^2 x} +
\frac{1}{(1-q)^4 x^2} \sum_{n=0}^{\infty}\frac{(-1)^n
A_n(q)}{(n+2)!(q-1)^{2n}}(\frac{1}{x})^n\right)
\]
follows from Lemma \ref{psr}.
\end{remark}

\section{$q$-Fourier transform for symmetric functions}
The {\it $q$-Fourier transform} for $q \ge 1$, based on the
$q$-product, was introduced in \cite{UmarovTsallisSteinberg} and
played a central role in establishing the $q$-analog of the standard
central limit theorem. Formally the $q$-Fourier transform
 for a given function $f(x)$ is defined by the
formula
\begin{equation}
\label{FourierTr}
F_q[f](\xi) = \int_{-\infty}^{\infty} e_q^{ix\xi} \otimes_q f(x) dx \,.
\end{equation}
For discrete functions $f_k, k=0, \pm 1,...,$ this definition takes the
form
\begin{equation}
\label{FourierDiscrete}
F_q[f](\xi) = \sum_{k= -\infty}^{\infty} e_q^{ik\xi} \otimes_q f(k) \,.
\end{equation}
In the future we use the same notation in both cases. We also call
(\ref{FourierTr}) or
(\ref{FourierDiscrete}) the {\it $q$-characteristic function} of a
given
random variable $X$ with
an associated density $f(x),$ using the
notations $F_q(X)$ or $F_q(f)$
equivalently.
\par
It should be noted that, if in the formal definition (\ref{FourierTr}),
$f$ is compactly
supported, then integration has to be taken over this support, although, in contrast with the usual analysis, the function
$e_q^{ix\xi} \otimes_q f(x)$ under the integral does not vanish outside the
support of $f$. This is an effect of
the $q$-product.
\par
The following lemma establishes the relation of the
$q$-Fourier transform without
using the $q$-product.
\begin{lem}
\label{informal}
The $q$-Fourier transform can be written in the form
\begin{equation}
\label{identity2} F_q[f](\xi) = \int_{-\infty}^{\infty}f(x)
e_q^{{ix\xi}{(f(x))^{q-1}}}dx.
\end{equation}
\end{lem}
\begin{remark}
\label{reminformal} Note that, if the $q$-Fourier transform of a
given function $f(x)$ defined by the formal definition in
(\ref{FourierTr}) exists, then it coincides with the expression in
(\ref{identity2}). The $q$-Fourier transform determined by the
formula (\ref{identity2}) has an advantage when compared to the
formal definition: it does not use the $q$-product, which is, as we
noticed above, restrictive in use.
\end{remark}
\par
Further to the properties of the $q$-Fourier transform established in \cite{UmarovTsallisSteinberg}, we note that, for
symmetric densities, the assertion analogous to Lemma \ref{informal} is true with the $q$-cos.
\begin{lem}
\label{cos2} Let $f(x)$ be an even function. Then its $q$-Fourier
transform can be written in the form
\begin{equation}
\label{identity3}
F_q[f](\xi) = \int_{-\infty}^{\infty}f(x) \cos_q ({x \xi
[f(x)]^{q-1}})dx.
\end{equation}
\end{lem}
\vspace{.3cm}

{\it Proof.}
Notice that, because of the symmetry of $f$,
\[
\int_{-\infty}^{\infty}e_q^{ix\xi} \otimes_q f(x)dx =
\int_{-\infty}^{\infty}e_q^{-ix \xi} \otimes_q f(x)dx  \,.
\]
Taking this into account, we have
\[
F_q[f](\xi) = \frac{1}{2} \int_{-\infty}^{\infty} \left( e_q^{ix\xi}
\otimes_q f(x) +  e_q^{- ix\xi} \otimes_q f(x) \right) dx \,.
\]
Applying Lemma \ref{informal} we obtain
\[
F_q[f](\xi) = \int_{-\infty}^{\infty} f(x) \frac{ e_q^{ix\xi
[f(x)]^{q-1}}  +  e_q^{- ix\xi [f(x)]^{q-1} }}{2}  dx \,,
\]
which coincides with (\ref{identity3}). \eproof

Further, denote $H_{q,\alpha}=\{f \in L_1: f(x) \sim C
|x|^{-\frac{1+\alpha}{1+\alpha(q-1)}}, \, \, |x|\rightarrow
\infty\}.$ It is readily seen that $\phi
(q,\alpha)=\frac{\alpha+1}{1+\alpha(q-1)}>1$ for all $\alpha \in
(0,2)$ and $q \in [1,2).$ Moreover, $\phi(q,\alpha)(2q-1)<3$ for all
$\alpha \in (0,2)$ and $q \in [1,2),$ which implies
$\sigma_{2q-1}^2(f)=\infty.$

\begin{lem}
\label{mainlemma} Let $f(x), \, x \in R, $ be a symmetric
probability density function of a given random variable. Further,
let either
\begin{itemize}
\item[(i)] the $(2q-1)$-variance $\sigma_{2q-1}^2 (f) <
\infty,$\footnote{In \cite{UmarovTsallisSteinberg} we did not
require the condition for $f(x)$ to be symmetric if $\sigma_{2q-1}^2
(f) < \infty.$} (associated with $(2q-1,\alpha) \in \mathcal{Q}_1$),
or
\item[(ii)]
$f(x) \in H_{q,\alpha}$, where $(2q-1, \alpha) \in \mathcal{Q}_2$.
\end{itemize}

Then, for the $q$-Fourier transform of $f(x)$, the following asymptotic
relation holds true:
\begin{equation}
\label{asforfi}
F_q[f](\xi)= 1 - \mu_{q,\alpha} |\xi|^{\alpha} + o(|\xi|^{\alpha}), \xi
\rightarrow 0,
\end{equation}
where
\begin{equation}
\mu_{q,\alpha} = \left\{ \begin{array}{ll}
  \vspace{1cm}
          \frac{q}{2} \sigma_{2q-1}^2 \nu_{2q-1} , &
          \mbox{if $\alpha = 2$ \,;} \\
          \   \frac{2^{2-\alpha}(1+\alpha(q-1))C}{2-q}  \int_0^{\infty}
\frac{- \, \Psi_q (y)}{y^{\alpha+1}} dy, & \mbox{if $(2q-1, \alpha)
\in \mathcal{Q}_2$ \,.}
  \end{array} \right.
\end{equation}
with $\nu_{2q-1}(f)= \int_{-\infty}^{\infty} [f(x)]^{2q-1}  \, dx$ .
\end{lem}
\begin{remark}
Stable distributions require $\mu_{q, \alpha}$ to be positive. We have seen (Lemma \ref{lempsi}) that if $q\ge1$, then
$\Psi_q(x) \le 0$ (not being identically zero), which yields $\mu_{q,\alpha} > 0 \,.$
\end{remark}
\vspace{.3cm}

{\it Proof.} First, we assume that $\alpha = 2$.  Using Lemma
\ref{informal} we have
\begin{equation}
\label{step_10}
F_q [f](\xi) = \int_{-\infty}^{\infty}(e_q^{ix \xi}) \otimes_q f(x) dx
=
                              \int_{-\infty}^{\infty} f(x)
\cos_q (x \xi [f(x)]^{q-1})x  \,.
\end{equation}
Making use of the asymptotic expansion (\ref{exp})
we can rewrite the right hand side of (\ref{step_10})
in the form
$$F_q [f](\xi) =  \\
   \int_{-\infty}^{\infty} f(x) \left(1 + i x \xi [f(x)]^{q-1}
- q/2 x^2 \xi^2 [f(x)]^{2(q-1)} \right)dx + o(\xi^3 )
 =$$
\[
  1  - (q/2) \xi^2 \sigma^{2}_{2q-1} \nu_{2q-1}+
o(\xi^3 ), \, \xi \rightarrow 0,
\]
from which the first part of Lemma follows.

Now, assume $(2q-1, \alpha) \in \mathcal{Q}_2.$ We apply Lemma
\ref{cos2} to obtain
\[
F_q[f](\xi)-1 = \int_{-\infty}^{\infty} f(x) [\cos_q ({x \xi
[f(x)]^{q-1}}) - 1] dx =
\]
\[
2 \int_{0}^{N} f(x) \Psi_q ( \frac {x \xi [f(x)]^{q-1}}{2})dx +
2 \int_{N}^{\infty} f(x) \Psi_q (\frac {x \xi [f(x)]^{q-1}}{2}) dx \,,
\]
where $N$ is a sufficiently large finite number. In the first integral we use the asymptotic relation
$\Psi({x \over 2})= - {q \over 2} x^2 + o(x^3)$, which follows from Lemma \ref{lempsi}, and get
\par
$
2 \int_{0}^{N} f(x) \Psi_q (\frac {x \xi [f(x)]^{q-1}}{2})dx =
$
\begin{equation}
\label{1integral}
-q \xi^2 \int_0^N x^2 f^{2q-1}(x)dx + o(\xi^3), \, \xi \rightarrow 0.
\end{equation}
In the second integral
taking into account the hypothesis of the lemma with respect to $f(x)$,
we have
\[
2 \int_{N}^{\infty} f(x) \Psi_q ( \frac {x \xi [f(x)]^{q-1}}{2}) dx
= 2C \int_{N}^{\infty} \frac{1}{x^{\frac{\alpha +
1}{1+\alpha(1-q)}}} \Psi_q ( \frac {x^{1-\frac{(\alpha
+1)(q-1)}{1+\alpha(q-1)}} \xi}{2} ) dx \,.
\]
We use the substitution
\[
x^{\frac{2-q}{1+\alpha(q-1)}} = \frac{2y}{\xi}
\]
in the last integral, and obtain
\[
2 \int_{N}^{\infty} f(x) \Psi_q (\frac{ x \xi [f(x)]^{q-1}}{2}) dx =
\]
\begin{equation}
\label{2integral} -\frac{2^{2-\alpha}(1+\alpha(q-1))C}{2-q}
|\xi|^{\alpha} \, \, \int_0^{\infty} \frac{\Psi_q (y)}{y^{\alpha+1}}
dy + o(|\xi|^{\alpha}), \, \, \xi \rightarrow 0.
\end{equation}
Hence, the obtained asymptotic relations (\ref{1integral}) and
(\ref{2integral}) complete the proof. \eproof \vspace{.3cm}

\section{Weak convergence of correlated random variables}

This section we start with introduction of the notion of {\it
$q$-independence} in particular case. More general definition and
some examples are given in \cite{UmarovTsallisSteinberg}. We will
also introduce two types of convergence, namely, {\it
$q$-convergence} and {\it weak $q$-convergence} and establish their
equivalence.

\begin{definition}
Two random variables $X$ and $Y$ are said to be $q$-independent
\footnote{We assume $X$ and $Y$ to have the zero $q$-means. For the
definition of $q$-independence for random variables with non-zero
$q$-means see \cite{UmarovTsallisSteinberg} }
 if
\begin{equation}
\label{q-correlation} F_q[X+Y](\xi)=F_q[X](\xi) \otimes_q
F_q[Y](\xi) \,.
\end{equation}
\end{definition}
In terms of densities, the relation (\ref{q-correlation}) can be
rewritten as follows. Let $f_X$ and $f_Y$ be densities of $X$ and
$Y$ respectively, and let $f_{X+Y}$ be the density of $X+Y$. Then
\begin{equation}
\label{q-correlation2} \int_{-\infty}^{\infty} e_q^{i x \xi}
\otimes_q f_{X+Y}(x) dx = F_q[f_X](\xi) \otimes_q F_q[f_Y](\xi).
\end{equation}
For $q=1$ the conditions (\ref{q-correlation}) turns into the well
known relation
\[
F[f_X \ast f_Y] = F[f_X] \cdot F[f_Y]
\]
between the convolution (noted $\ast$) of two densities and the
multiplication of their (classical) Fourier images, and holds for
independent $X$ and $Y$. If $q \neq 1$, then $q$-independence
describes a specific correlation.
\begin{definition}
Let $X_1, X_2,...,X_N,...$ be a sequence of identically distributed
random variables. Denote $Y_N=X_1+...+X_N$. By definition, the
sequence $X_N, N=1,2,...$ is said to be $q$-independent (or
$q$-i.i.d.) if for all $N=2,3,...$, the relations
\begin{equation}
\label{qiid1} F_{q}[Y_N ](\xi)=F_{q}[X_1](\xi) \otimes_{q} ...
\otimes_{q} F_{q}[X_N](\xi)
\end{equation}
hold.
\end{definition}
\begin{definition}
\label{qconvergence} A sequence of random variables $X_N, \,
N=1,2,...,$ is said to be $q$-convergent to a random variable
$X_{\infty}$ if $\lim_{N \rightarrow \infty} F_q [X_N](\xi) = F_q
[X_{\infty}](\xi)$ locally uniformly in $\xi$.
\end{definition}

Evidently, this definition is equivalent to the weak convergence
(denoted by "$\Rightarrow$") of random variables if $q=1.$ For $q
\neq 1$ denote by $W_q$ the set of continuous functions $\phi$
satisfying the condition $|\phi(x)| \leq C(1+|x|)^{-\frac{q}{q-1}},
\, x \in R$.

\begin{definition}
\label{qweakconvergence} A sequence of random variables $X_N$ with
the density $f_N$ is called weakly $q$-convergent to a random
variable $X_{\infty}$ with the density $f$ if $\int_{R} f_N(x) d m_q
\rightarrow \int_{R^d} f(x)d m_q $ for arbitrary measure $m_q$
defined as $dm_q(x)= \phi_q (x) dx,$ where $\phi_q \in W_q$. We
denote the $q$-convergence by the symbol
$\stackrel{q}{\Rightarrow}$.
\end{definition}

\begin{lem}
\label{weak} Let $q>1.$ Then $X_N \Rightarrow X_0 $ yields $X_N
\stackrel{q}{\Rightarrow} X_0$.
\end{lem}

The proof of this lemma immediately follows from the obvious fact
that $W_q$ is a subset of the set of bounded continuous functions.

Recall that a sequence of
probability measures $\mu_N$ is called tight if, for an arbitrary
$\epsilon >0$, there is a compact $K_{\epsilon}$ and an integer
$N^{\ast}_{\epsilon}$ such that $\mu_N(R^d \setminus K_{\epsilon}) <
\epsilon$ for all $N \geq N_{\epsilon}^{\ast}$.

\begin{lem}
\label{tight} Let $1<q<2.$ Assume a sequence of random variables
$X_N$, defined on a probability space with a probability measure
$P$, and associated densities $f_N,$ is $q$-convergent to a random
variable $X$ with an associated density $f.$ Then the sequence of
associated probability measures $\mu_N = P(X_N^{-1})$ is tight.
\end{lem}

{\it Proof.}
Assume that $1<q<2$ and $X_N$ is a $q$-convergent sequence of random
variables with associated densities $f_N$ and associated probability
measures $\mu_N$. We have
\[
\frac{1}{R}\int_{-R}^{R}(1-F_q[f_N](\xi)) d \xi =
\frac{1}{R}\int_{-R}^{R}(1-\int_{R} f_N e_q^{i x \xi f_N^{q-1}} dx )
d \xi =
\]
\begin{equation}
\label{tight1} \int_{R} \left( \frac{1}{R}\int_{-R}^{R}(1- e_q^{i x
\xi f_N^{q-1}} ) d \xi \right) d \mu_N(x).
\end{equation}
It is not hard to verify that
\begin{equation}
\label{tight2} \frac{1}{R}\int_{-R}^{R} e_q^{i x \xi t} d \xi =
\frac{2 \sin_{\frac{1}{2-q}} (R x(2-q) t)}{Rx(2-q)t}.
\end{equation}
It follows from (\ref{tight1}) and (\ref{tight2}) that
\begin{equation}
\label{tight3} \frac{1}{R}\int_{-R}^{R}(1-F_q[f_N](\xi)) d \xi =
2\int_{-\infty}^{\infty} \left( 1 - \frac{\sin_{ \frac{1}{2-q} }
(x(2-q)R f_N^{q-1})}{Rx(2-q)f_N^{q-1}} \right) d\mu_N(x).
\end{equation}
Since $1<q<2$ by assumption,  $\frac{1}{2-q} > 1$ as well. It is
known \cite{qborges1,qnivanen,qborges2} that for any $q^{'}>1$ the
properties $sin_{q^{'}}(x) \leq 1$ and $(sin_{q^{'}}(x))/x
\rightarrow 1, \, x \rightarrow 0$ hold. Moreover,
$(\sin_{q^{'}}(x))/x \leq 1, \forall x \in R.$ Suppose,
$lim_{|x|\rightarrow \infty} |x|f_N^{q-1}=L_N, \, N \ge 1.$ Divide
the set $\{N \ge N_0\}$ into two subsets $A=\{N_j \ge N_0: L_{N_j} >
1\}$ and $B=\{N_k \ge N_0: L_{N_k} \le 1\}.$
If $N \in A,$ since $\sin_{\frac{1}{2-q}} \le 1,$ there is a number
$a>0$ such that
\[
\frac{1}{R}\int_{-R}^{R}(1-F_q[f_N](\xi)) d \xi \geq 2 \int_{|x|
\geq a} \left ( 1-\frac{1}{R|x|(2-q)f_N^{q-1}} \right) d\mu_N (x)
\]
\[
 \geq C \mu_N \left( |x| \geq a
\right), \, \, C>0 \, ~~ \forall \, N \in A,
\]
for $R$ small enough. Now taking into account the $q$-convergence of
$X_N$ to $X$ and, if necessary, taking $R$ smaller, for any
$\epsilon
>0$, we obtain
\[
\mu_N \left( |x| \geq a \right) \leq \frac{1}{C
R}\int_{-R}^{R}(1-F_q[f_0](\xi)) d \xi < \epsilon, \,  \, ~~ \forall
\, N \in A.
\]
If $N \in B$
then there exist constants $b>0, \, \delta>0,$ such that
\[
f_N(x) \le \frac{L_N+\delta}{|x|^{\frac{1}{q-1}}} \le
\frac{1+\delta}{|x|^{\frac{1}{q-1}}}, \, |x|\ge b, \forall \, N \in
B.
\]
Hence, we have
\[
\mu_N(|x|>b) = \int_{|x|>b}f_N(x)dx \le {(1+\delta)}
\int_{|x|>b}\frac{dx}{|x|^{\frac{1}{q-1}}}, \, N \in B.
\]
Since, $1/(q-1)>1,$ for any $\epsilon >0$ we can select a number
$b_{\epsilon} \ge b$ such that $\mu_N(|x|>b_{\epsilon})<\epsilon, \,
N \in B.$ As far as $A \cup B = \{N \ge N_0\}$ the proof of the
statement is complete. \eproof

Further, we introduce  the function
\begin{equation}
\label{dt} D_q(t)= D_q(t; a)=t e_q^{i a t^{q-1}}= t(1 + i (1-q) a
t^{q-1})^{-\frac{1}{q-1}},
\end{equation}
defined on  $ [0,1]$, where $1<q<2$ and $a$ is a fixed real number.
Obviously $D_q(t)$ is continuous on $[0,1]$ and differentiable in
the interval $(0,1)$. In accordance with the classical Lagrange
average theorem for any $t_1, t_2, \, \, 0 \leq t_1 < t_2 \leq 1$
there exists a number $t_{\ast}, \, \, t_1 < t_{\ast} < t_2$ such
that
\begin{equation}
\label{ev} D_q(t_1)-D_q(t_2) = D_q^{'}(t_{\ast}) (t_1 - t_2),
\end{equation}
where $D_q^{'}$ means the derivative of $D_q(t)$ with respect to
$t$.

Consider the following Cauchy problems for the Bernoulli equation
\begin{equation}
\label{bern1} y^{'} - \frac{1}{t} y = \frac{ia(1-q)}{t} y^q, \, \,
y(0)=0,
\end{equation}
It is not hard to verify that $D_q(t)$ is a solution to the problems
(\ref{bern1}).

\begin{lem}
For $D^{'}_q (t)$  the estimate
\begin{equation}
\label{est} |D_q^{'}(t;a)| \leq C (1 + |a|)^{-\frac{q}{q-1}}, \, \,
t \in (0,1], \, a \in R^1,
\end{equation}
holds, where the constant $C$ does not depend on $t$.
\end{lem}

{\it Proof.} It follows from (\ref{dt}) and (\ref{bern1}) that
\[
|y^{'}(t)| \leq t^{-1}|y + i a (1-q)y^{q}| = |e_q^{i a t^{q-1}} + i
a (1-q)(e_q^{i a t^{q-1}})^q|=
\]
\[
|1+ia(1-q)t^{q-1}|^{-\frac{q}{q-1}}  \leq C
(1+|a|)^{-\frac{q}{q-1}}, \, t \in (0,1].
\]
\eproof
\begin{thm}
\label{continuitytheorem1} Let $1<q<2$ and a sequence of random
vectors $X_N$ be weakly $q$-convergent to a random vector $X$. Then
$X_N$ is $q$-convergent to $X$.
\end{thm}

{\it Proof.}
%Let $B_{r_k}$ is a ball in $R^d$ with a radius $r_k, \, \, r_k \rightarrow \infty, \, k \rightarrow \infty$.
Assume $X_N,$ with associated densities $f_N,$ is weakly
$q$-convergent to a $X,$ with an associated density $f$. The
difference $\mathcal{F}_q [f_N](\xi) - \mathcal{F}_q [f_N](\xi)$ can
be written in the form
\begin{equation}
\label{difference} \mathcal{F}_q [f_N](\xi) - \mathcal{F}_q
[f_N](\xi)
       =  \int_{R^d} \left( D_q(f_N(x)) - D_q( f(x) ) \right) dx,
\end{equation}
where $D_q(t)=D_q(t;a)$ is defined in (\ref{dt}) with $a = x \xi$.
It follows from (\ref{ev}) and (\ref{est}) that
\[
|\mathcal{F}_q [f_N](\xi) - \mathcal{F}_q [f_N](\xi)| \leq C
\int_{R^d} | (1+|x|)^{-\frac{q}{q-1}} \left(f_N(x)- f(x) \right)|dx,
\]
which yields $\mathcal{F}_q [f_N](\xi) \rightarrow \mathcal{F}_q
[f_N](\xi)$ for all $\xi \in R^d$. \eproof
\begin{thm}
Let $1<q<2$ and a sequence of random vectors $X_N$ with the
associated densities $f_N$ is $q$-convergent to a random vector $X$
with the associated density $f$ and $\mathcal{F}_q[f](\xi)$ is
continuous at $\xi = 0$. Then $X_N$ weakly $q$-converges to $X$.
\end{thm}

 {\it Proof.} Now assume that $f_N$ converges to $f$ in the sense of
$q$-convergence. It follows from Lemma \ref{tight} that the
corresponding sequence of induced probability measures $\mu_N =
P(X_N^{-1})$ is tight. This yields relatively weak compactness of
the sequence $\mu_N.$ Theorem \ref{continuitytheorem1} implies that
each weakly convergent subsequence $\{\mu_{N_j}\}$ of $\mu_N$
converges to $\mu=P(X^{-1}).$ Hence, $\mu_N \Rightarrow \mu$, or the
same, $X_N \Rightarrow X.$ Now applying Lemma \ref{weak} we complete
the proof. \eproof

\section{Symmetric $(q,\alpha)$-stable distributions. First representation}

In this section we introduce symmetric $(q,\alpha)$-stable
distributions and give the description based on the mapping
(\ref{paper2part1}). In accordance with this description $q$ takes
any value in $[1,2),$ however we distinguish the cases $\alpha = 2$
and $\alpha \in (0,2).$
\begin{definition}
\label{stabledistr} A random variable $X$ is said to have a
$(q,\alpha)$-stable distribution if its $q$-Fourier transform is
represented in the form $e_q^{-\beta |\xi|^{\alpha}}$, with $\beta
>0.$
 We denote by $\mathcal{L}_q(\alpha)$ the set of all $(q,\alpha)$-stable distributions.
\end{definition}
Denote $ \mathcal{G}_q(\alpha) = \{b \, e_q^{-\beta |\xi|^{\alpha}},
\, \, b > 0, \, \, \beta >0\}.$ In other words $X \in
\mathcal{L}_q(\alpha)$ if $F_q[f] \in \mathcal{G}_q(\alpha)$ with
$b=1.$ Note that if $\alpha = 2$, then $\mathcal{G}_q(2)$ represents
the set of $q$-Gaussians and $\mathcal{L}_q(2)$ - the set of random
variables whose densities are $q_{\ast}$-Gaussians, where
$q_{\ast}=(3q-1)/(1+q)$.
\begin{prop}
\label{q-stability} Let $q$-independent random variables $X_j \in
\mathcal{L}_q(\alpha), j=1,..,m.$ Then for any constants
$a_1,...,a_m,$
\[
\sum_{j=1}^m a_j X_j \in \mathcal{L}_q(\alpha).
\]
\end{prop}

{\it Proof.} Let $$F_q[X_j](\xi)=e_q^{-\beta_j}|\xi|^{\alpha}, \,
j=1,...,m.$$ Using the properties $e_q^x \otimes_q e_q^y =e_q^{x+y}$
and $F_q[aX](\xi)=F_q[X](a^{2-q}\xi),$ it follows from the
definition of the $q$-independence that
\[
F_q[\sum_{j=1}^m a_j X_j] = e_q^{-\beta |\xi|^{\alpha}}, \, \beta =
\sum_{j=1}^m \beta_j |a|^{\alpha(2-q)} > 0.
\]
\eproof
\begin{remark} Proposition \ref{q-stability} justifies the stability
of distributions in $\mathcal{L}_q(\alpha).$ Recall that if $q=1$
then $q$-independent random variables are independent in the usual
sense. Thus, if $q=1, \, 0<\alpha<2,$ then $\mathcal{L}_1(\alpha)$
coincides with symmetric $\alpha$-stable L\'evy distributions
$\mathcal{L}_{sym}(\alpha)$.
\end{remark}
Further we show that the $q$-weak limits of sums
\[
Z_N = \frac{1}{s_{N}(q,\alpha)} \, (X_1 + ...+ X_N ), N=1,2,...
\]
as $N \rightarrow \infty,$ where $s_{N}(q,\alpha), \, N=1,2,...,$
are some reals (scaling parameter), also belong to
$\mathcal{L}_q(\alpha).$
\begin{definition}
A sequence of random variables $Z_N$ is said to be $q$-convergent to
a $(q,\alpha)$-stable distribution, if $\lim_{N\rightarrow \infty}
F_q[Z_N](\xi) \in \mathcal{G}_q(\alpha)$ locally uniformly by $\xi$.
\end{definition}

{\bf Theorem 1.} {\it Assume $(2q-1,\alpha) \in \mathcal{Q}_2.$ Let
$X_1,X_2,...,X_N,...$ be symmetric $q$-independent random variables
and all having the same probability density function $f(x) \in
H_{q,\alpha}.$ Then $Z_N$, with $s_N(q,\alpha) =
(\mu_{q,\alpha}N)^{\frac{1}{\alpha (2-q)}}$, is $q$-convergent to a
$(q,\alpha)$-stable distribution, as $N \rightarrow \infty.$}
\par
\begin{remark}
By definition $\mathcal{Q}_2$ excludes the value $\alpha=2$. The
case $\alpha=2$, in accordance with the first part of Lemma
\ref{mainlemma}, coincides with Theorem 2 of
\cite{UmarovTsallisSteinberg}. Note in this case $\mathcal{L}_q(2)=
\mathcal{G}_{q^{\ast}}(2)$, where $q^{\ast}=\frac{3q-1}{q+1}.$
\end{remark}
\par
{\it Proof.} Assume $(Q,\alpha) \in \mathcal{Q}_2.$ Let $f$ be the
density associated with $X_1$. First we evaluate
$F_q(X_1)=F_q(f(x)).$ Using Lemma \ref{mainlemma} we have
\begin{equation}
\label{step_1}
F_q[f](\xi)= 1 - \mu_{q,\alpha} |\xi|^{\alpha} + o(|\xi|^{\alpha}), \xi
\rightarrow 0.
\end{equation}
Denote $Y_j = N^{-\frac{1}{\alpha}} \, X_j, j=1,2,...$. Then $Z_N =
Y_1 +...+Y_N.$ Further, it is readily seen that, for a given random
variable $X$ and real $a>0$, the equality $F_q
[aX](\xi)=F_q[X](a^{2-q} \xi)$ holds. It follows from this relation
that $F_q(Y_j)=F_q[f]( \frac{\xi}{ (\mu_{q,\alpha}N)^{1/\alpha } }
), \, j=1,2,...$ Moreover, it follows from the $q$-independence of
$X_1,X_2,...,$ and the associativity of the $q$-product that
\begin{equation}
\label{step2}
F_q[Z_N](\xi)= F_q[f]( (\mu_{q,\alpha}N)^{-\frac{1}{\alpha} }\xi )
{\otimes_q ... \otimes_q}
F_q[f]( (\mu_{q,\alpha}N)^{-\frac{1}{\alpha}}\xi )
\,\,(N\,\mbox{factors}).
\end{equation}
Hence, making use of the expansion (\ref{log}) for the $q$-logarithm, Eq. (\ref{step2}) implies
\[
\ln_q F_q[Z_N](\xi)= N \ln_q F_q[f](
(\mu_{q,\alpha}N)^{-\frac{1}{\alpha}}\xi ) =
N \ln_q ( 1- \frac{|\xi|^{\alpha}}{N} + o(\frac{|\xi|^{\alpha}}{N})) =
\]
\begin{equation}
\label{step101}
- |\xi|^{\alpha} + o(1), \, N \rightarrow \infty,
\end{equation}
locally uniformly by $\xi$.
\par
Hence, locally uniformly by $\xi,$
\begin{equation}
\label{step_3}
\lim_{N \rightarrow \infty} F_q(Z_N) = e_q^{- |\xi|^{\alpha}} \in
\mathcal{G}_q(\alpha).
\end{equation}
Thus, $Z_N$ is $q$-convergent to a $(q,\alpha)$-stable distribution,
as $N \rightarrow \infty.$ \eproof

\begin{figure}[hpt]
\centering
\includegraphics[width=3.5in]{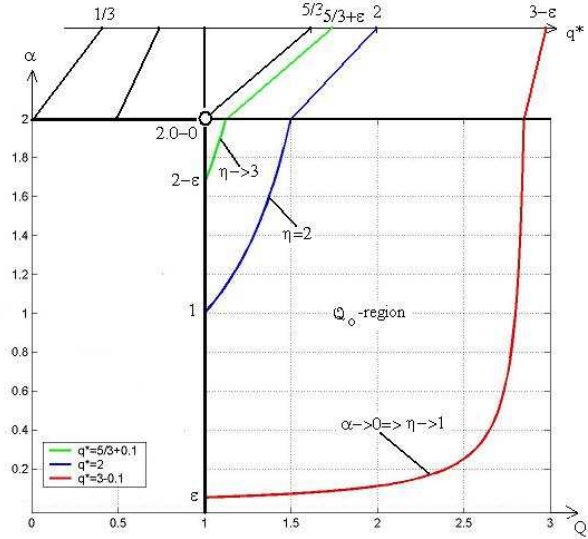}
\caption{\scriptsize All pairs of $(Q,\alpha)$ on the indicated
curves are associated with the same $q^L_{\alpha}$-Gaussian. Two
curves corresponding to two different values of $q_{\alpha}^L$ do
not intersect. In this sense these curves represent  the constant
levels of $q_{\alpha}^L$ or $\eta=2/(q_{\alpha}^L - 1).$ The line
$\eta=1$ joins the points $(Q,\alpha)=(1,0.0-0)$ and $(3-0,2)$; the
line  $\eta=2$ joins the Cauchy distribution (noted $C$) with itself
at $(Q,\alpha)=(1,1)$ and at $(2,2)$; the $\eta=3$ line joins the
points $(Q,\alpha)=(1,2.0-0)$ and $(5/3,2)$ (by $\epsilon$ we simply
mean to give an indication, and not that both infinitesimals
coincide). The entire line at $Q=1$ and $0<\alpha<2$ is mapped into
the line at $\alpha=2$ and $5/3 \le q^{\ast}<3$. } \label{fig:Alpha}
\end{figure}

\begin{figure}[hpt]
\centering
\includegraphics[width=2.5in,clip=]{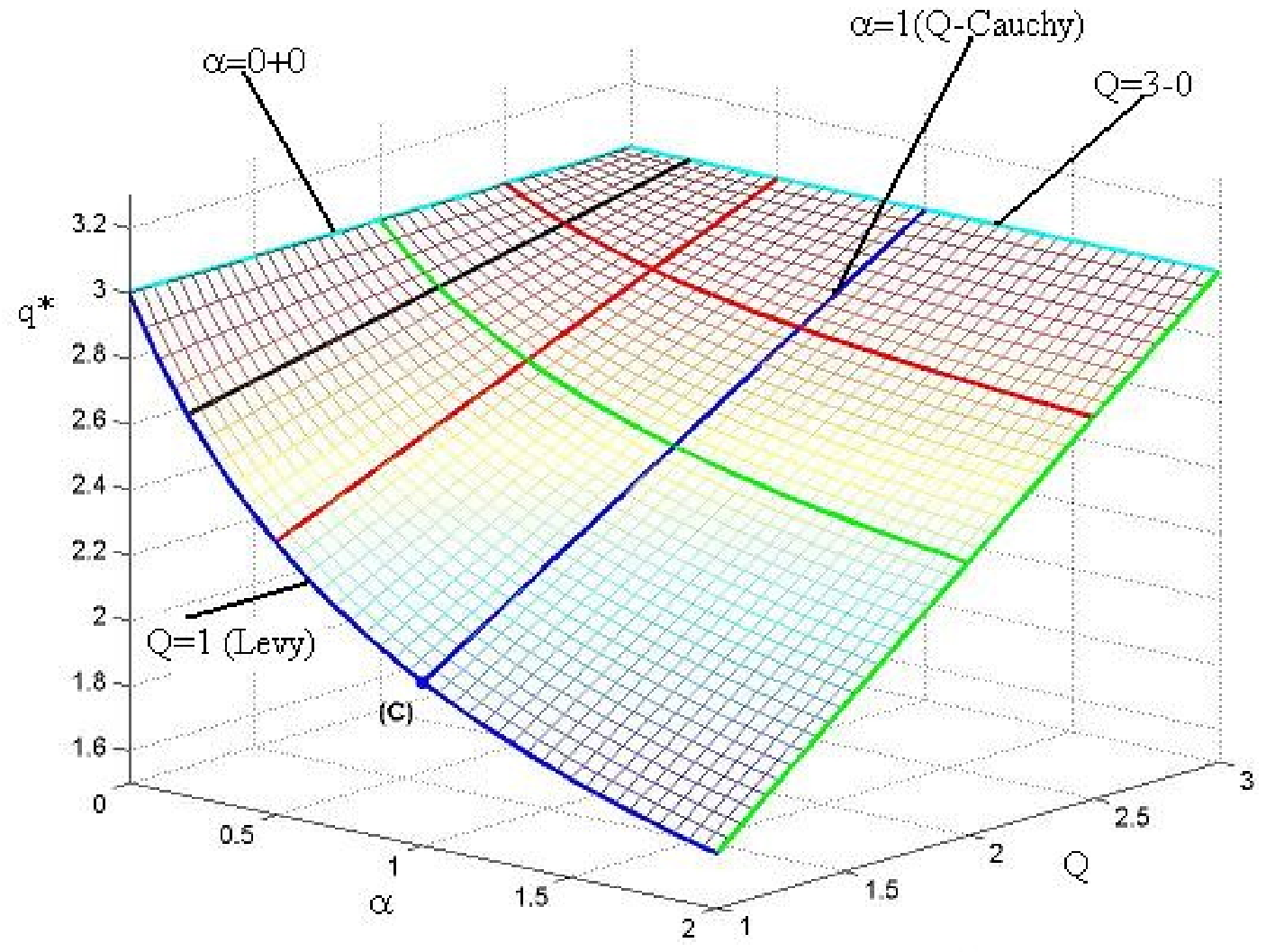}
\includegraphics[width=2.5in]{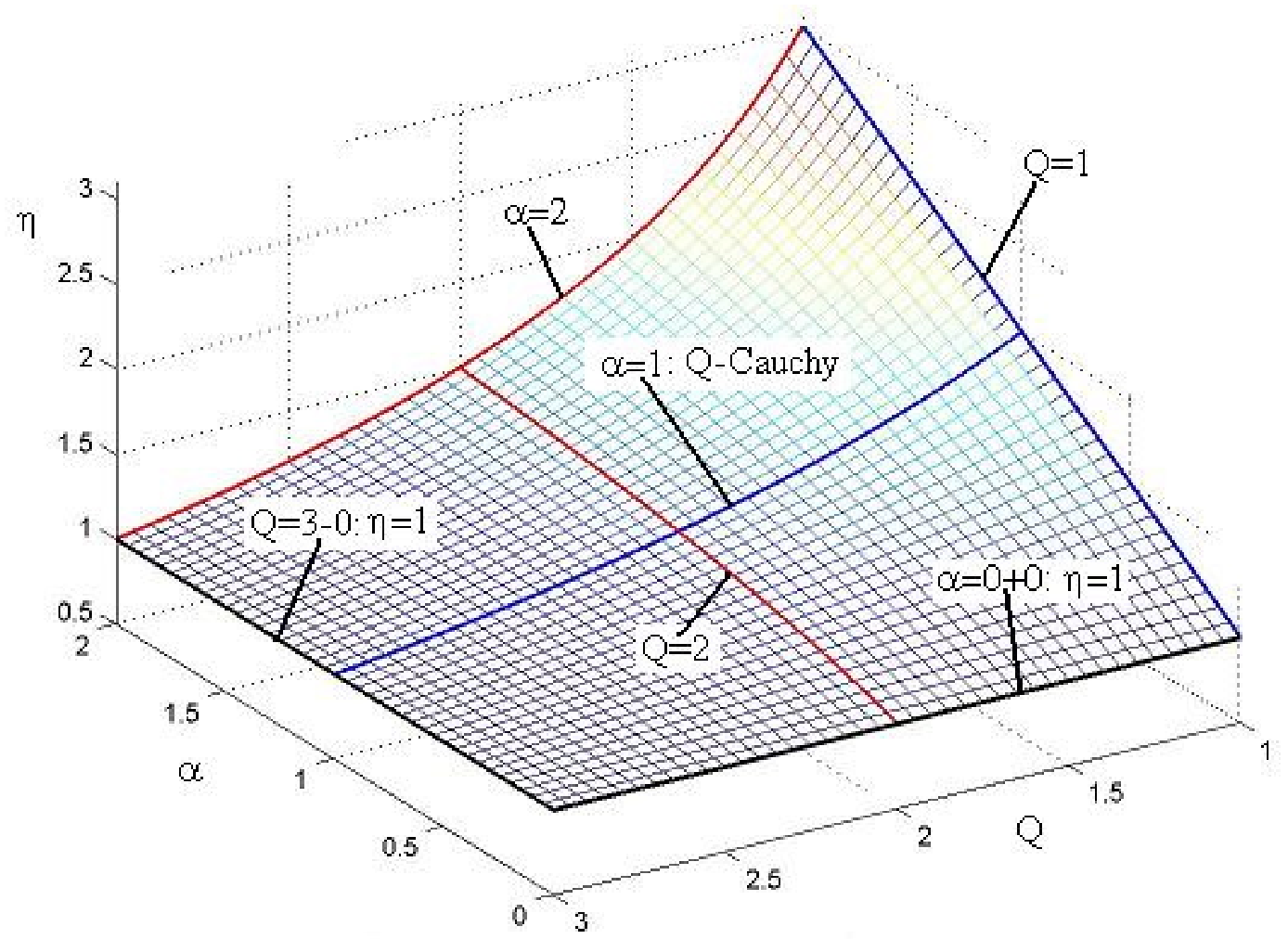}
\caption{\scriptsize $q^L_{\alpha}\equiv q^*$ and $\eta$ as
functions of variables $(Q,\alpha).$ } \label{fig:Eta}
\end{figure}

This theorem links the classic L\'evy
distributions with
their $q^{L}_{\alpha}$-Gaussian counterparts. Indeed, in accordance with this theorem, a function $f$, for which
\[
f \sim C/  x^{ (\alpha +1)/(1 + \alpha (q-1)) }, \, \, |x| \rightarrow
\infty,
\]
is in $\mathcal{L}_q(\alpha),$ i.e. $F_q[f](\xi) \in
\mathcal{G}_q(\alpha).$ It is not hard to verify that there exists a
$q^{L}_{\alpha}$-Gaussian, which is asymptotically equivalent to
$f$. Let us now find $q^{L}_{\alpha}.$ Any $q^{L}_{\alpha}$-Gaussian
behaves asymptotically $C_1/|x|^{\eta} = C_2/ |x|^{
2/(q^{L}_{\alpha}-1) }, \, C_j=const, \, j=1,2$, i.e.
$\eta=2/(q^{L}_{\alpha}-1).$ Hence, we obtain the relation
\begin{equation}
\frac{\alpha +1}{1 + \alpha (q-1) } = \frac{2}{q^{L}_{\alpha}-1}.
\end{equation}
Solving this equation with respect to $q_{\alpha}^L$, we have
\begin{equation}
\label{general}
q^{L}_{\alpha} = \frac{3+ Q \alpha }{\alpha + 1}, \, \, Q=2q-1 \,,
\end{equation}
linking three parameters: $\alpha,$ the parameter of the
$\alpha$-stable L\'evy distributions,  $q,$ the parameter of
correlation, and $q^{L}_{\alpha}$, the parameter of attractors in
terms of $q^{L}_{\alpha}$-Gaussians (see Fig. \ref{fig:Eta} (left)).
Equation (\ref{general}) identifies all $(Q,\alpha)$-stable
distributions with the same index of attractor $G_{q^{L}_{\alpha}}$
(See Fig. \ref{fig:Alpha}), proving the following proposition.

\begin{prop}
Let $1\le Q <3, \, (Q=2q-1)$ and $0<\alpha<2.$ Then all
distributions $X \in \mathcal{L}_q(\alpha),$ where the pairs
$(Q,\alpha)$ satisfy the equation
\[
\frac{3+Q \alpha}{\alpha+1}=q_{\alpha}^L,
\]
have the same attractor asymptotically equivalent to
$q_{\alpha}^L$-Gaussian.
\end{prop}

\begin{figure}[hpt]
\centering
\includegraphics[width=2.5in]{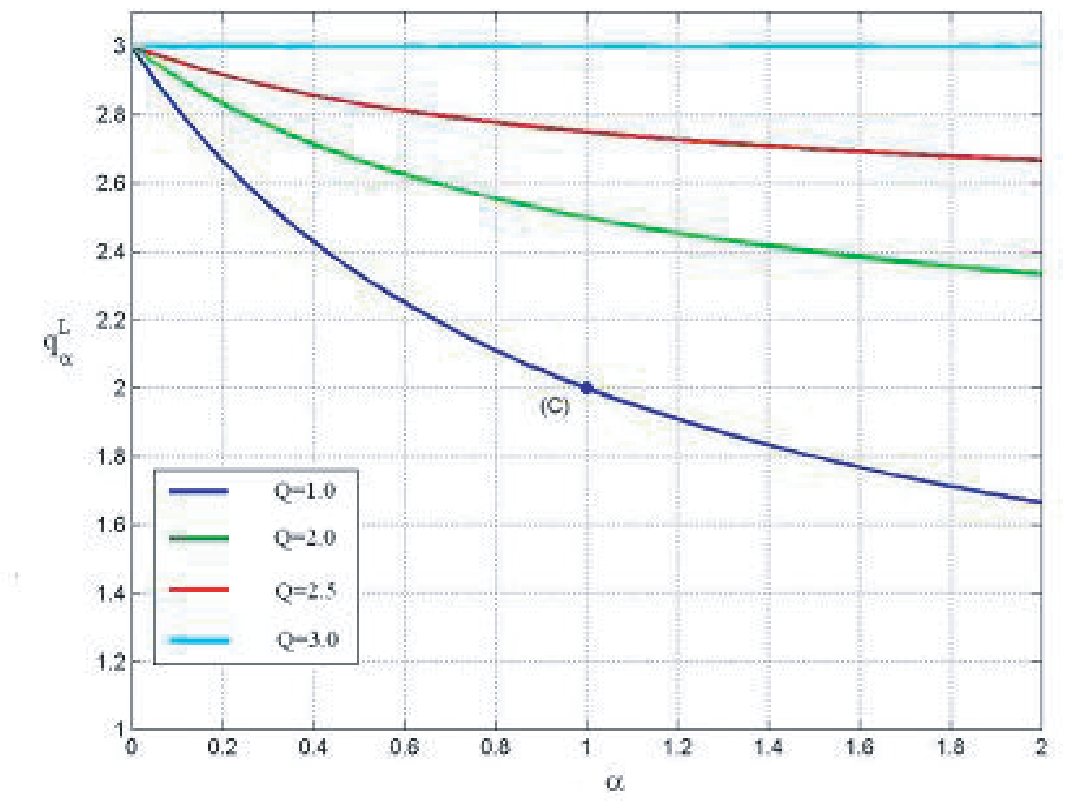}
\includegraphics[width=2.5in]{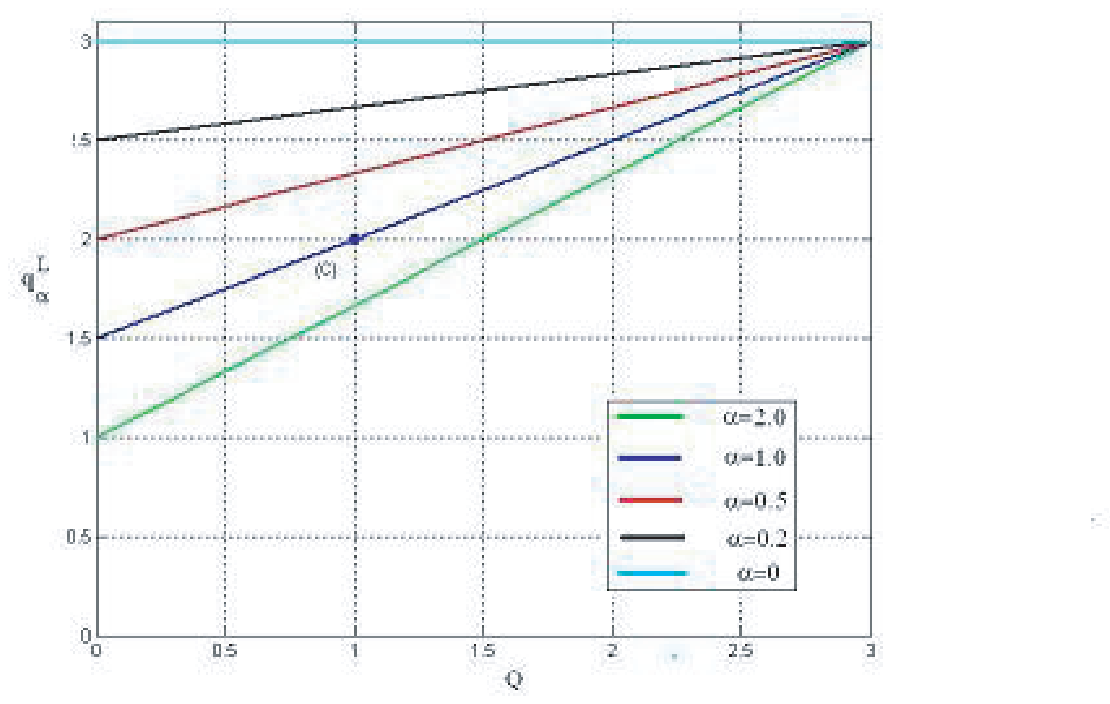}\\
\caption{\scriptsize Constant $Q$ and constant $\alpha$ sections of
Fig. \ref{fig:Eta}(left). } \label{fig:Fig4}
\end{figure}

In the particular case $Q=1$, we recover the known connection
between the classical L\'evy distributions ($q = Q = 1$) and
corresponding $q^{L}_{\alpha}$-Gaussians. Put $Q=1$ in Eq.
(\ref{general}) to obtain
\begin{equation}
\label{Q=1}
q^{L}_{\alpha} = \frac{3+\alpha}{1+\alpha}, \, \, 0<\alpha<2.
\end{equation}
When $\alpha$ increases between $0$ and $2$ (i.e. $0<\alpha<2$),
$q^{L}_{\alpha}$ decreases between $3$ and $5/3$ (i.e. $5/3 <
q^L_{\alpha}<3$): See Figs. \ref{fig:Eta} (left) and
\ref{fig:Fig4}(left).

It is useful to find the relationship between $\eta =
\frac{2}{q^{L}_{\alpha}-1}$, which corresponds to the asymptotic
behaviour of the attractor depending on $(\alpha,Q)$. Using formula
(\ref{general}), we obtain (Fig. \ref{fig:Eta}, right)
\begin{equation}
\label{eta}
\eta = \frac{2(\alpha + 1)}{2 + \alpha (Q-1)}.
\end{equation}

\begin{prop}
Let $X \in \mathcal{L}_Q(\alpha), \, 1\le Q <3, \, 0<\alpha<2.$ Then
the associated density function $f_X$ has asymptotics $f_X(x)\sim
|x|^{\eta}, \, |x|\rightarrow \infty,$ where $\eta=\eta(Q,\alpha)$
is defined in (\ref{eta}).
\end{prop}

\begin{remark}
If $Q=1$ (classic L\'evy distributions), then $\eta = \alpha + 1$,
as is well known.
\end{remark}

Analogous relationships can be obtained  for other values of $Q$. We
call, for convenience, a $(Q,\alpha)$-stable distribution  a
$Q$-Cauchy distribution, if its parameter $\alpha = 1.$ We obtain
the classic Cauchy-Poisson distribution if  $Q = 1$. The
corresponding line can be obtained cutting the surface in Fig.
\ref{fig:Eta} (right)  along the line $\alpha = 1$. For $Q$-Cauchy
distributions we have
\begin{equation}
\label{Q-Cauchy}
q^{L}_{1}(Q) = \frac{3+Q}{2} \, \, \, \mbox{and} \, \, \, \eta=\frac{4}{Q+1},
\end{equation}
respectively (see Figs. \ref{fig:Eta}).

The relationship between $\alpha$ and $q^{L}_{\alpha}$ for typical
fixed values of $Q$ are given in Fig. \ref{fig:Fig4} (left). In this
figure we can also see, that $\alpha = 1$ (Cauchy)  corresponds to
$q^{L}_{1}=2$ (in the $Q=1$ curve). In Fig. \ref{fig:Fig4} (right)
the relationships between $Q$ ($Q=2q-1$) and $q^{L}_{\alpha}$ are
represented for typical fixed values of  $\alpha$.

We acknowledge thoughtful remarks by R.
Hersh, E.P. Borges and S.M.D. Queiros. We thank E. Andries for his valuable assistance with MatLab codes and figures.
Financial support by the Fullbright Foundation, SI International,
AFRL and NIH grant P20 GMO67594 (USA agencies), and CNPq, Pronex and Faperj (Brazilian agencies) are acknowledged as well.

\end{document}